\documentclass[aps, prl, twocolumn, superscriptaddress, showpacs]{revtex4}

\usepackage{mathrsfs}
\usepackage[dvips]{graphicx}
\begin{document}
\title
{
  Deviation from the Cosmological Constant or Systematic Errors?
}

\author{Shi Qi}
\email{qishi11@gmail.com}
\affiliation
{
  Purple Mountain Observatory, Chinese Academy of Sciences, Nanjing
  210008, China
}
\affiliation
{
  Joint Center for Particle, Nuclear Physics and Cosmology, Nanjing
  University -- Purple Mountain Observatory, Nanjing  210093, China
}
\affiliation
{
  Kavli Institute for Theoretical Physics China, CAS, Beijing 100190,
  China
}

\author{Tan Lu}
\email{t.lu@pmo.ac.cn}
\affiliation
{
  Purple Mountain Observatory, Chinese Academy of Sciences, Nanjing
  210008, China
}
\affiliation
{
  Joint Center for Particle, Nuclear Physics and Cosmology, Nanjing
  University -- Purple Mountain Observatory, Nanjing  210093, China
}

\author{Fa-Yin Wang}
\email{fayinwang@nju.edu.cn}
\affiliation
{
  Department of Astronomy, Nanjing University, Nanjing 210093, China
}

\date{\today}

\begin{abstract}
  Motivated by the fact that both SNe Ia and GRBs seem to prefer a
  dark energy EOS greater than $-1$ at redshifts $z \gtrsim 0.5$, we
  perform a careful investigation on this situation.
  We find that the deviation of dark energy from the cosmological
  constant at redshifts $z \gtrsim 0.5$ is large enough that we should
  pay close attention to it with future observational data.
  Such a deviation may arise from some biasing systematic errors in
  the handling of SNe Ia and/or GRBs or more interestingly from the
  nature of the dark energy itself.
\end{abstract}

\pacs{95.36.+x, 98.80.Es}


\maketitle

It has been ten years since the discovery of the cosmic
acceleration~\cite{Riess:1998cb, Perlmutter:1998np}, which is
attributed to the mysterious component --- dark energy.
In addition to the cosmological constant, a lot of dark energy models
have been proposed to explain the cosmic acceleration (see for
example~\cite{Copeland:2006wr}).
Though the standard $\Lambda$CDM model fits the observational data
well, there are also a variety of other dark energy models could not
be ruled out due to the precision of current data.
It is therefore still a crucial issue that whether the dark energy is
simply the cosmological constant or not.

Among all kinds of observational sources, type Ia supernova (SN Ia),
which has been widely used as standard candles, is one of the most
important classes of data that could impose significant constraints on
the nature of dark energy.
One of the important reasons for this is that SNe Ia provide data
points along the redshifts and could therefore recover the nature of
dark energy at different redshifts.
However, due to the limitation of the redshifts of SNe Ia, it is
difficult to study the nature of dark energy beyond redshift of $1.7$
with SNe Ia.
While gamma-ray bursts (GRBs), as the most luminous astrophysical
events observed today, extend the redshift to $z>6$.
After calibrated, they could be used as complementary sources to SNe
Ia at high redshifts in cosmology studies
(see e.g.~\cite{Schaefer:2006pa}),
which has recently attracted much attention.
At present, GRBs are still not as ideal standard candles as SNe Ia.
The scatters of known luminosity relations of GRBs are still very
large and they have circularity problem due to the lack of low
redshift samples.
In spite of this, works by many authors have put it forward in their
cosmic applications.
For example, recent advances include that new luminosity relations are
introduced~\cite{Dainotti:2008vw, Tsutsui:2008sy}
and that model-independent calibrations
are proposed~\cite{Li:2007re, Qi:2008zk, Kodama:2008dq, Liang:2008kx,
  Wang:2008vja}.
Among works using combined data of SNe Ia and GRBs, it is notable
in~\cite{Qi:2008zk} that the improvements on the constraints made by
including GRBs show that the dark energy equation
of state (EOS) is slightly shifted towards $w > -1$
at redshifts $z \gtrsim 0.5$~\footnote{In the
  published version of~\cite{Qi:2008zk}, there are typos in the
  luminosity relations Eq.~(16) and Eq.~(17) and the typos are passed
  on to Eq.~(21). However, the correct equations have been used in the
  calculations, so the results are unaffected. The typos have been
  corrected in third version on arXiv.}.
Since the results there are still totally consistent with the
cosmological constant at $2 \sigma$ confidence level and the inclusion
of GRBs is much preliminary (Only systematic errors of luminosity
relations are included for simplicity), we cannot draw any concrete
conclusion only from that.
However, it is interesting that, when we have more SN Ia samples, SNe
Ia themselves also show the same trend,
as was shown by Figure~17 in~\cite{Kowalski:2008ez}.
(To be fair, There are also other analyses with earlier SN Ia sets as
well as with the Union set~\cite{Kowalski:2008ez} show the sign of the
possible increase in the dark energy EOS, see for
example~\cite{Alam:2003fg, Alam:2006kj, Sahni:2008xx}.
The advantages of the results in~\cite{Qi:2008zk} and Figure~17
in~\cite{Kowalski:2008ez} are that the redshift binned
parameterization is used, which assumes less about the nature of the
dark energy compared to simple parameterizations especially at high
redshifts.
See, for example, in~\cite{Sullivan:2007pd}, where the redshift binned
parameterization are adopted with earlier SN Ia sets, the constraints
on the dark energy EOS at redshifts $z \gtrsim 0.5$ are still too
weak. See also Figure~11 in~\cite{Riess:2006fw} for an illustration of
priors imposed on the dark energy by a simple parameterization
itself.)
Because GRBs and SNe Ia are independent sources, the fact that they
both seem to prefer a dark energy EOS greater than $-1$ at redshifts
$z \gtrsim 0.5$ may be worth our more attention.
Motivated by this, we perform a careful investigation on this
situation in this paper.

In~\cite{Qi:2008zk}, five luminosity relations are used for GRBs,
i.e.
\begin{eqnarray}
  \label{eq:GRB-lag-L}
  \log \frac{L}{1 \; \mathrm{erg} \; \mathrm{s}^{-1}}
  &=& a_1+b_1 \log
  \left[
    \frac{\tau_{\mathrm{lag}}(1+z)^{-1}}{0.1\;\mathrm{s}}
  \right]
  ,
  \\
  \label{eq:GRB-V-L}
  \log \frac{L}{1 \; \mathrm{erg} \; \mathrm{s}^{-1}}
  &=& a_2+b_2 \log
  \left[
    \frac{V(1+z)}{0.02}
  \right]
  ,
  \\
  \label{eq:GRB-E_peak-L}
  \log \frac{L}{1 \; \mathrm{erg} \; \mathrm{s}^{-1}}
  &=& a_3+b_3 \log
  \left[
    \frac{E_{\mathrm{peak}}(1+z)}{300\;\mathrm{keV}}
  \right]
  ,
  \\
  \label{eq:GRB-E_peak-E_gamma}
  \log \frac{E_{\gamma}}{1\;\mathrm{erg}}
  &=& a_4+b_4 \log
  \left[
    \frac{E_{\mathrm{peak}}(1+z)}{300\;\mathrm{keV}}
  \right]
  ,
  \\
  \label{eq:GRB-tau_RT-L}
  \log \frac{L}{1 \; \mathrm{erg} \; \mathrm{s}^{-1}}
  &=& a_5+b_5 \log
  \left[
    \frac{\tau_{\mathrm{RT}}(1+z)^{-1}}{0.1\;\mathrm{s}}
  \right]
  .
\end{eqnarray}
To avoid circularity problem, calibration parameters and cosmological
parameters are fitted simultaneously, and in the calculation of
$\chi^2$, only the systematic errors of the luminosity relations are
taken into account for simplicity in the preliminary study of the
evolution of the dark energy EOS including GRBs
(see~\cite{Qi:2008zk} for details).
In this paper, we include also the measurement errors of GRBs for a
careful investigation, i.e.,
$\sigma_{\mathrm{tot}}^2 = \sigma_{\mathrm{mea}}^2 +
\sigma_{\mathrm{sys}}^2$,
where the systematic errors, $\sigma_{\mathrm{sys}}$s, are derived by
requiring the reduced $\chi^2$ equal to $1$
and the measurement errors, $\sigma_{\mathrm{mea}}$s, are given by
$\sigma_{\mathrm{mea}}^2 = \sigma_y^2 + b^2 \sigma_x^2$
for the fitting to
\begin{equation}
  \label{eq:fit}
  y = a + b x
  ,
\end{equation}
where $x$ and $y$ denote the logarithm of the luminosity indicators
and the logarithm of luminosity or energy of GRBs
(see Eq.~(\ref{eq:GRB-lag-L}-\ref{eq:GRB-tau_RT-L})).
For asymmetric measurement errors, the errors of the side near the
line being fitted to are used~\cite{Wang:2008vja}.

In addition to GRBs, we have used Union compilation of SNe Ia
from~\cite{Kowalski:2008ez}, BAO measurement
from~\cite{Eisenstein:2005su} and $\Omega_m h = 0.213 \pm 0.023$
from~\cite{Tegmark:2003uf}.
We assumed the prior $\Omega_k=-0.014 \pm 0.017$~\cite{Spergel:2006hy}
for the cosmic curvature.
We adopted the redshift binned parameterization for the dark energy
EOS, as proposed in~\cite{Huterer:2004ch}, to estimate possible
evolution of the dark energy.
In this parameterization, the redshifts are divided into several bins
and the dark energy EOS is taken to be constant in each redshift bin
but can vary from bin to bin.
And a set of decorrelated EOS parameters are introduced subsequently
by an appropriate transformation.
The evolution of the dark energy with respect to the redshift could be
estimated from these decorrelated EOS parameters.
In this paper, we divided redshifts at points $z =0.2,\ 0.5,\ 1$ and
Markov chain Monte Carlo techniques are used with $\mathscr{O}(10^6)$
samples generated for each result.
Since current observational data have only very weak constraints on
the nature of dark energy at redshifts $z > 1$, we focus our analyses
on the first three redshift bins.

\begin{figure}[tbp]
  \centering
  \includegraphics[width = 0.45 \textwidth]{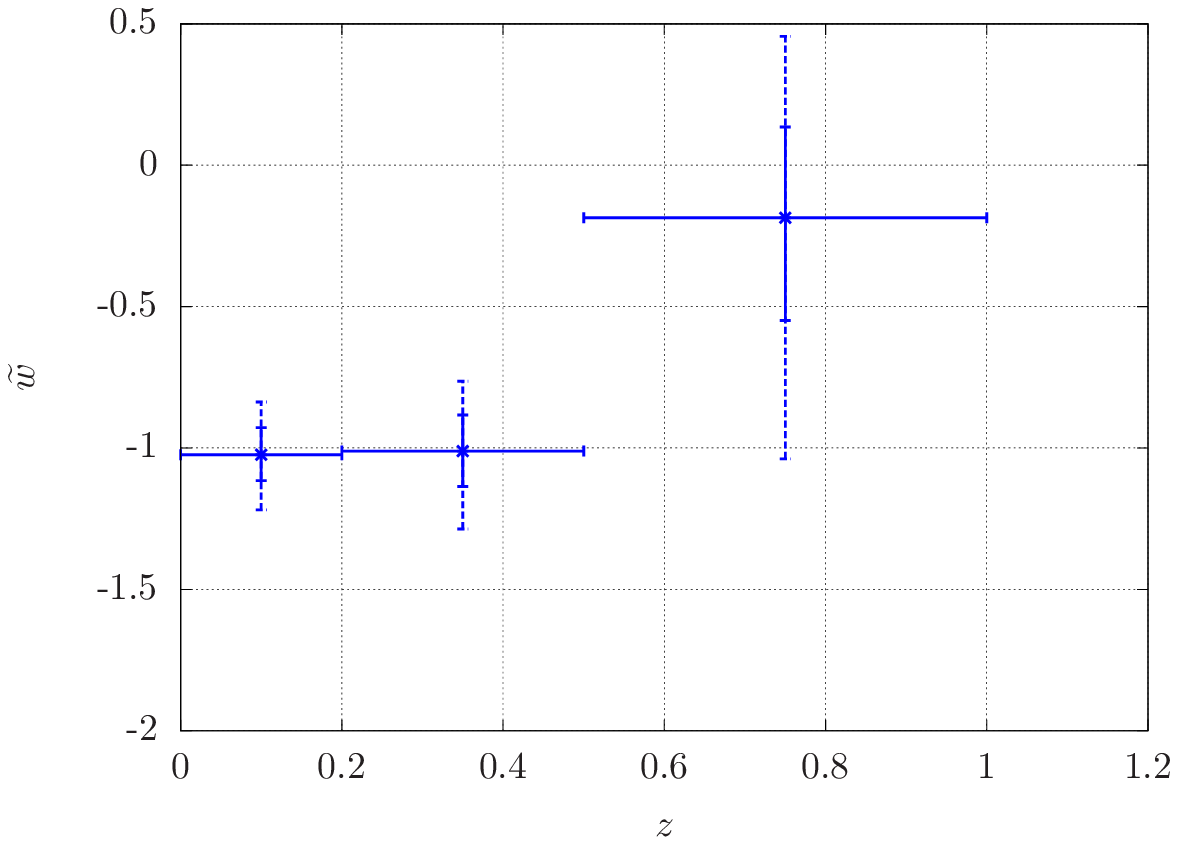}
  \includegraphics[width = 0.45 \textwidth]{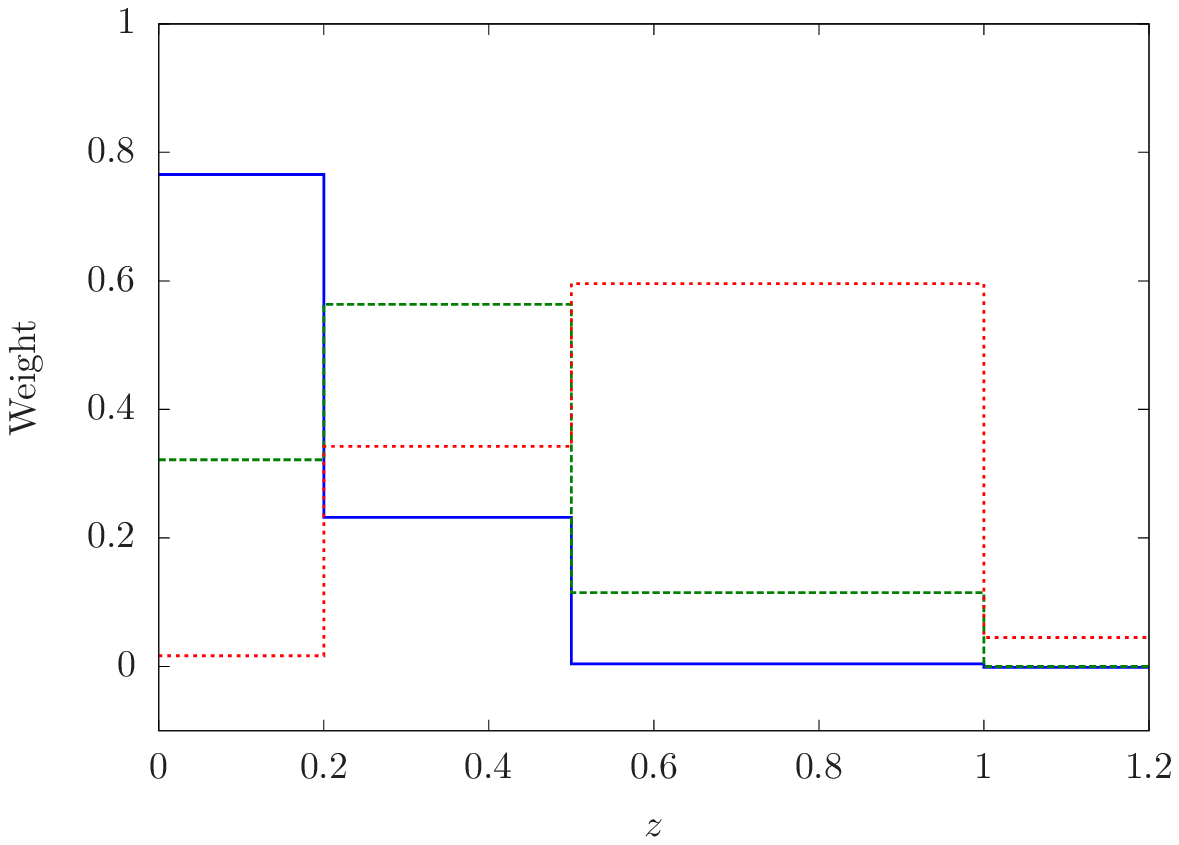}
  \caption
  {
    Estimates of the uncorrelated dark energy EOS parameters
    $\widetilde{w}_i$.
    Top: uncorrelated dark energy parameters versus redshift, in which
    the vertical errorbars correspond to $1 \sigma$ and $2 \sigma$
    confidence levels of $\widetilde{w}_i$ and the horizontal
    errorbars span the corresponding redshift bins from which the
    contributions to $\widetilde{w}_i$ come most.
    Bottom: Window functions for $\widetilde{w}_i$.
  }
  \label{fig:wz_and_weight}
\end{figure}

\begin{figure}[tbp]
  \centering
  \includegraphics[width = 0.45 \textwidth]{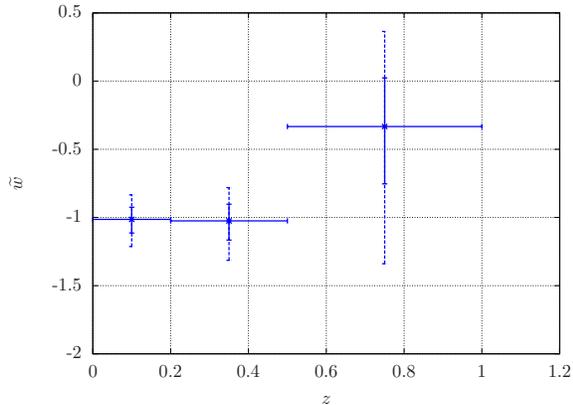}
  \caption
  {
    Estimates of the uncorrelated dark energy EOS parameters. Same as
    the top plot of Figure~\ref{fig:wz_and_weight} except that GRBs
    are not included in constraining.
  }
  \label{fig:wz_without_GRB}
\end{figure}

Figure~\ref{fig:wz_and_weight} shows the results derived from data set
including GRBs.
As stated above, calibration parameters of GRBs and cosmological
parameters are fitted simultaneously, and measurement errors and
systematic errors are both taken into account.
We can see that the deviation from the cosmological constant at
redshifts $z \gtrsim 0.5$ turns out to be greater than the results
in~\cite{Qi:2008zk}, such that the EOS of $-1$ lies almost at the edge
of the $2 \sigma$ confidence interval.
Though still consistent with the cosmological constant at $2 \sigma$
confidence level, such a deviation should be large enough to attract
our attention.
Of course, if this deviation is just an illustration of statistical
errors due to the limitation of current observational data, it would
be meaningless and should disappear with the increase of the
observational data.
While a comparison of the top plot of
Figure~\ref{fig:wz_and_weight} with Figure~\ref{fig:wz_without_GRB},
for which GRBs are not included in constraining, shows that SNe Ia
alone shift the dark energy EOS at redshifts $z \gtrsim 0.5$ upwards
from the cosmological constant and GRBs shift it a little more in the
same direction.
This means that both SNe Ia and GRBs prefer a dark energy EOS greater
than $-1$ at redshifts $z \gtrsim 0.5$.
One can argue that the independence of SNe Ia and GRBs reduces the
possibility of the deviation arising from statistical errors.
Such a deviation from the cosmological constant, if confirmed,
may be caused by the nature of the dark energy itself or some biasing
systematic errors in the observational data that should be excluded.
For the latter, we would need reconsider the process of calibrating
SNe Ia and/or GRBs.
It is notable that the recent CfA3 addition of SN Ia samples has
brought SN Ia cosmology to the point where systematic uncertainties
dominate~\cite{Hicken:2009df, Hicken:2009dk}.
While the former is more exciting for possibly ruling out the
cosmological constant as the dark energy.
A close attention should be paid to this deviation with future
observational data.

\begin{figure}[tbp]
  \centering
  \includegraphics[width = 0.45 \textwidth]{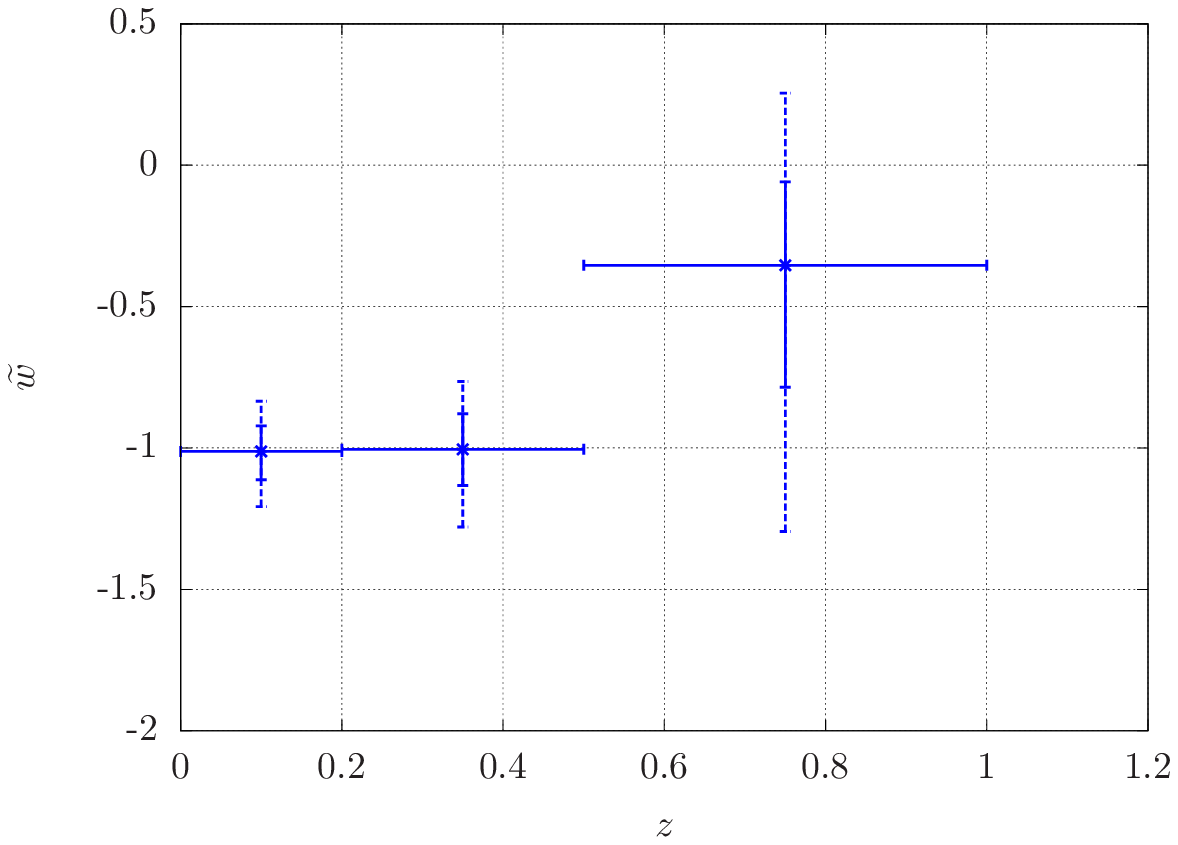}
  \includegraphics[width = 0.45 \textwidth]{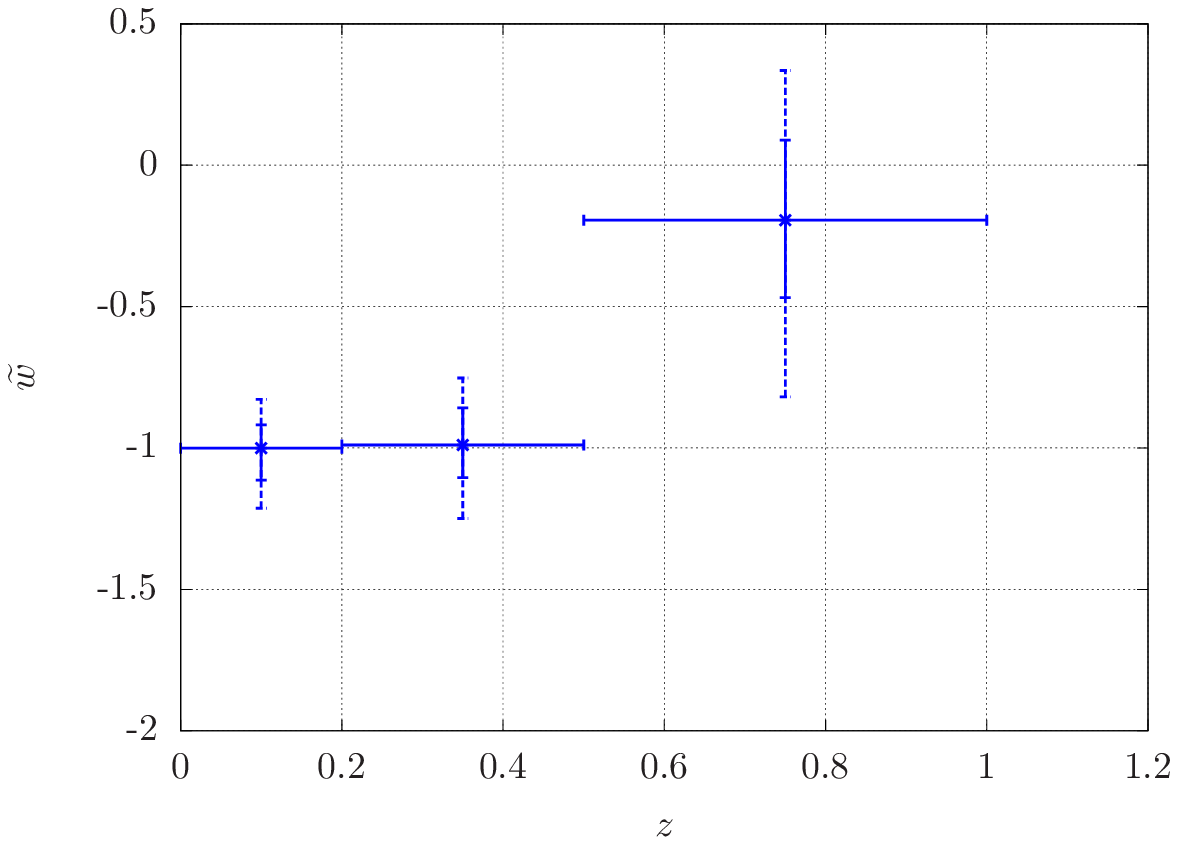}
  \caption
  {
    Estimates of the uncorrelated dark energy EOS parameters. Same as
    the top plot of Figure~\ref{fig:wz_and_weight} except that GRBs
    are included in model-independent ways.
    Top: GRBs are included by using the distance measurements
    from~\cite{Wang:2008vja}.
    Bottom: GRBs are included by using calibrated GRBs of redshifts
    $z > 1.4$ from~\cite{Cardone:2009mr}.
  }
  \label{fig:wz_mi}
\end{figure}

To be more careful, we crosschecked our results by including GRBs in
other different ways.
For the results in Figure~\ref{fig:wz_and_weight}, GRB data are not
processed prior to being used to constrain the dark energy, i.e. the
calibration of GRBs and constraining cosmological parameters are
carried out simultaneously.
In light of model-independent calibrations of GRBs in
literatures~\cite{Kodama:2008dq, Liang:2008kx, Wang:2008vja},
we performed the same analyses including instead the preprocessed
GRB data from~\cite{Wang:2008vja} and~\cite{Cardone:2009mr}.
In~\cite{Wang:2008vja}, GRB data are summarized by a set of
model-independent distance measurements.
These distance measurements can be used directly to replace GRBs in
constraining cosmological parameters.
In~\cite{Cardone:2009mr}, GRBs of redshift $z \leq 1.4$ are utilized
to calibrate the luminosity relations based on a local regression
estimate of distance moduli using the Union SN Ia
sample~\cite{Kowalski:2008ez},
so the GRBs of redshifts $z > 1.4$, whose distance moduli are derived
from the calibrated luminosity relations, can be used in the same way
as SNe Ia.
We present in Figure~\ref{fig:wz_mi} the results of including GRBs in
the above two ways.
The bottom plot of Figure~\ref{fig:wz_mi},
for which the calibrated GRBs of redshifts $z > 1.4$
from~\cite{Cardone:2009mr} are used,
is consistent with the results in
Figure~\ref{fig:wz_and_weight}, except that the constraints are
slightly tighter such that the cosmological constant has been ruled
out at $2 \sigma$ confidence level at redshifts $z \gtrsim 0.5$,
which can be easily understood --- some of the SNe Ia are used in both
calibrating GRBs and constraining cosmological parameters.
However, the top plot of Figure~\ref{fig:wz_mi}, for which GRBs are
included by
using the distance measurements from~\cite{Wang:2008vja}, is somewhat
different from the results in Figure~\ref{fig:wz_and_weight}.
Comparing with the results without including GRBs
(see Figure~\ref{fig:wz_without_GRB}),
we can see that including these distance measurements does not change
the result much.
In fact, it was shown in~\cite{Wang:2008vja} that these distance
measurements shift best-fit parameter values towards the cosmological
constant.
Since the derivation of the distance measurements from GRBs involves
quite a few intermediate steps and are carried out through the Markov
chain Monte Carlo method, it is obscure what has caused the
difference.

\begin{figure}[tbp]
  \centering
  \includegraphics[width = 0.45 \textwidth]{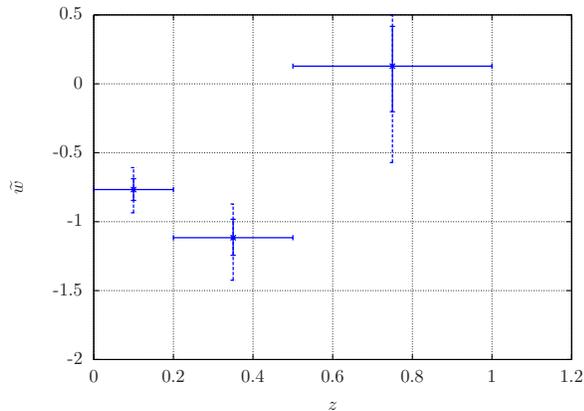}
  \caption
  {
    Estimates of the uncorrelated dark energy EOS parameters. Same as
    the top plot of Figure~\ref{fig:wz_and_weight} except that
    Constitution set of SNe Ia from~\cite{Hicken:2009dk} and the BAO
    measurements from~\cite{Percival:2007yw} are used instead.
  }
  \label{fig:wz_hp}
\end{figure}

At the end, we would like to mention that in the above analyses we did
not use the recent Constitution set of SNe Ia~\cite{Hicken:2009dk} and
the BAO measurements presented in~\cite{Percival:2007yw}.
First, this is for the consistency of the data.
Because, in~\cite{Cardone:2009mr}, GRBs are calibrated with the Union
set of SNe Ia~\cite{Kowalski:2008ez}.
Second, there seem to be some tension in these data sets.
The results derived using them are quite different from the above.
For the BAO measurements presented in~\cite{Percival:2007yw}, see the
argument in~\cite{Kowalski:2008ez}.
For SNe Ia, we noted that Union set prefers a Hubble parameter around
$70$ km/s/Mpc. While Constitution set are derived by adding CfA3 SNe
Ia to Union set using a Hubble parameter of $65$ km/s/Mpc.
We wonder whether this will cause any problems of consistency or not.
Anyway, we present in Figure~\ref{fig:wz_hp} the results using
Constitution set of SNe Ia~\cite{Hicken:2009dk} and the BAO
measurements from~\cite{Percival:2007yw}, leaving the clarification of
the differences between the data sets for the future.
In spite of this, we can see from Figure~\ref{fig:wz_hp} that our
conclusion on the dark energy EOS at redshifts $z \gtrsim 0.5$ are
unaffected.
See~\cite{Shafieloo:2009ti} for a discussion on the behaviour of the
dark energy at low redshifts derived from Constitution set of SNe
Ia~\cite{Hicken:2009dk} and the BAO measurements presented
in~\cite{Percival:2007yw}.

In summary,
motivated by the fact that both SNe Ia and GRBs seem to prefer a dark
energy EOS greater than $-1$ at redshifts $z \gtrsim 0.5$, we perform
a careful investigation on this situation, including more careful
treatments of measurement errors of GRBs than previous studies on the
evolution of the dark energy and crosscheck by using GRBs in different
ways.
We find that the deviation of dark energy from the cosmological
constant at redshifts $z \gtrsim 0.5$ is large enough that we should
pay close attention to it with future observational data.
Such a deviation may arise from some biasing systematic errors in the
handling of SNe Ia and/or GRBs or more interestingly from the nature
of the dark energy itself.

\begin{acknowledgments}
  This research was supported in part by the Project of Knowledge
  Innovation Program (PKIP) of Chinese Academy of Sciences, Grant
  No.~KJCX2.YW.W10 and the National Natural Science Foundation of
  China, Grant No.~10473023.
  Fa-Yin Wang was supported by the Jiangsu Project Innovation for PhD
  Candidates (CX07B-039z).
\end{acknowledgments}

\bibliography{dark_energy}

\begin{thebibliography}{26}
\expandafter\ifx\csname natexlab\endcsname\relax\def\natexlab#1{#1}\fi
\expandafter\ifx\csname bibnamefont\endcsname\relax
  \def\bibnamefont#1{#1}\fi
\expandafter\ifx\csname bibfnamefont\endcsname\relax
  \def\bibfnamefont#1{#1}\fi
\expandafter\ifx\csname citenamefont\endcsname\relax
  \def\citenamefont#1{#1}\fi
\expandafter\ifx\csname url\endcsname\relax
  \def\url#1{\texttt{#1}}\fi
\expandafter\ifx\csname urlprefix\endcsname\relax\def\urlprefix{URL }\fi
\providecommand{\bibinfo}[2]{#2}
\providecommand{\eprint}[2][]{\url{#2}}

\bibitem[{\citenamefont{Riess et~al.}(1998)}]{Riess:1998cb}
\bibinfo{author}{\bibfnamefont{A.~G.} \bibnamefont{Riess}} \bibnamefont{et~al.}
  (\bibinfo{collaboration}{Supernova Search Team}), \bibinfo{journal}{Astron.
  J.} \textbf{\bibinfo{volume}{116}}, \bibinfo{pages}{1009}
  (\bibinfo{year}{1998}), \eprint{astro-ph/9805201}.

\bibitem[{\citenamefont{Perlmutter et~al.}(1999)}]{Perlmutter:1998np}
\bibinfo{author}{\bibfnamefont{S.}~\bibnamefont{Perlmutter}}
  \bibnamefont{et~al.} (\bibinfo{collaboration}{Supernova Cosmology Project}),
  \bibinfo{journal}{Astrophys. J.} \textbf{\bibinfo{volume}{517}},
  \bibinfo{pages}{565} (\bibinfo{year}{1999}), \eprint{astro-ph/9812133}.

\bibitem[{\citenamefont{Copeland et~al.}(2006)\citenamefont{Copeland, Sami, and
  Tsujikawa}}]{Copeland:2006wr}
\bibinfo{author}{\bibfnamefont{E.~J.} \bibnamefont{Copeland}},
  \bibinfo{author}{\bibfnamefont{M.}~\bibnamefont{Sami}}, \bibnamefont{and}
  \bibinfo{author}{\bibfnamefont{S.}~\bibnamefont{Tsujikawa}},
  \bibinfo{journal}{Int. J. Mod. Phys.} \textbf{\bibinfo{volume}{D15}},
  \bibinfo{pages}{1753} (\bibinfo{year}{2006}), \eprint{hep-th/0603057}.

\bibitem[{\citenamefont{Schaefer}(2007)}]{Schaefer:2006pa}
\bibinfo{author}{\bibfnamefont{B.~E.} \bibnamefont{Schaefer}},
  \bibinfo{journal}{Astrophys. J.} \textbf{\bibinfo{volume}{660}},
  \bibinfo{pages}{16} (\bibinfo{year}{2007}), \eprint{astro-ph/0612285}.

\bibitem[{\citenamefont{Dainotti et~al.}(2008)\citenamefont{Dainotti, Cardone,
  and Capozziello}}]{Dainotti:2008vw}
\bibinfo{author}{\bibfnamefont{M.~G.} \bibnamefont{Dainotti}},
  \bibinfo{author}{\bibfnamefont{V.~F.} \bibnamefont{Cardone}},
  \bibnamefont{and}
  \bibinfo{author}{\bibfnamefont{S.}~\bibnamefont{Capozziello}},
  \bibinfo{journal}{MNRAS} \textbf{\bibinfo{volume}{391}}, \bibinfo{pages}{L79}
  (\bibinfo{year}{2008}), \eprint{0809.1389}.

\bibitem[{\citenamefont{Tsutsui et~al.}(2008)}]{Tsutsui:2008sy}
\bibinfo{author}{\bibfnamefont{R.}~\bibnamefont{Tsutsui}} \bibnamefont{et~al.}
  (\bibinfo{year}{2008}), \eprint{0810.1870}.

\bibitem[{\citenamefont{Li et~al.}(2008)}]{Li:2007re}
\bibinfo{author}{\bibfnamefont{H.}~\bibnamefont{Li}} \bibnamefont{et~al.},
  \bibinfo{journal}{Astrophys. J.} \textbf{\bibinfo{volume}{680}},
  \bibinfo{pages}{92} (\bibinfo{year}{2008}), \eprint{0711.1792}.

\bibitem[{\citenamefont{Qi et~al.}(2008)\citenamefont{Qi, Wang, and
  Lu}}]{Qi:2008zk}
\bibinfo{author}{\bibfnamefont{S.}~\bibnamefont{Qi}},
  \bibinfo{author}{\bibfnamefont{F.-Y.} \bibnamefont{Wang}}, \bibnamefont{and}
  \bibinfo{author}{\bibfnamefont{T.}~\bibnamefont{Lu}},
  \bibinfo{journal}{Astron. Astrophys.} \textbf{\bibinfo{volume}{483}},
  \bibinfo{pages}{49} (\bibinfo{year}{2008}), \eprint{arXiv:0803.4304
  [astro-ph]}.

\bibitem[{\citenamefont{Kodama et~al.}(2008)}]{Kodama:2008dq}
\bibinfo{author}{\bibfnamefont{Y.}~\bibnamefont{Kodama}} \bibnamefont{et~al.},
  \bibinfo{journal}{MNRAS} pp. \bibinfo{pages}{L1--L4} (\bibinfo{year}{2008}),
  \eprint{0802.3428}.

\bibitem[{\citenamefont{Liang et~al.}(2008)\citenamefont{Liang, Xiao, Liu, and
  Zhang}}]{Liang:2008kx}
\bibinfo{author}{\bibfnamefont{N.}~\bibnamefont{Liang}},
  \bibinfo{author}{\bibfnamefont{W.~K.} \bibnamefont{Xiao}},
  \bibinfo{author}{\bibfnamefont{Y.}~\bibnamefont{Liu}}, \bibnamefont{and}
  \bibinfo{author}{\bibfnamefont{S.~N.} \bibnamefont{Zhang}},
  \bibinfo{journal}{Astrophys. J.} \textbf{\bibinfo{volume}{685}},
  \bibinfo{pages}{354} (\bibinfo{year}{2008}), \eprint{0802.4262}.

\bibitem[{\citenamefont{Wang}(2008)}]{Wang:2008vja}
\bibinfo{author}{\bibfnamefont{Y.}~\bibnamefont{Wang}}, \bibinfo{journal}{Phys.
  Rev.} \textbf{\bibinfo{volume}{D78}}, \bibinfo{pages}{123532}
  (\bibinfo{year}{2008}), \eprint{0809.0657}.

\bibitem[{\citenamefont{Kowalski et~al.}(2008)}]{Kowalski:2008ez}
\bibinfo{author}{\bibfnamefont{M.}~\bibnamefont{Kowalski}}
  \bibnamefont{et~al.}, \bibinfo{journal}{Astrophys. J.}
  \textbf{\bibinfo{volume}{686}}, \bibinfo{pages}{749} (\bibinfo{year}{2008}),
  \eprint{0804.4142}.

\bibitem[{\citenamefont{Alam et~al.}(2004)\citenamefont{Alam, Sahni, Saini, and
  Starobinsky}}]{Alam:2003fg}
\bibinfo{author}{\bibfnamefont{U.}~\bibnamefont{Alam}},
  \bibinfo{author}{\bibfnamefont{V.}~\bibnamefont{Sahni}},
  \bibinfo{author}{\bibfnamefont{T.~D.} \bibnamefont{Saini}}, \bibnamefont{and}
  \bibinfo{author}{\bibfnamefont{A.~A.} \bibnamefont{Starobinsky}},
  \bibinfo{journal}{Mon. Not. Roy. Astron. Soc.}
  \textbf{\bibinfo{volume}{354}}, \bibinfo{pages}{275} (\bibinfo{year}{2004}),
  \eprint{astro-ph/0311364}.

\bibitem[{\citenamefont{Alam et~al.}(2007)\citenamefont{Alam, Sahni, and
  Starobinsky}}]{Alam:2006kj}
\bibinfo{author}{\bibfnamefont{U.}~\bibnamefont{Alam}},
  \bibinfo{author}{\bibfnamefont{V.}~\bibnamefont{Sahni}}, \bibnamefont{and}
  \bibinfo{author}{\bibfnamefont{A.~A.} \bibnamefont{Starobinsky}},
  \bibinfo{journal}{JCAP} \textbf{\bibinfo{volume}{0702}}, \bibinfo{pages}{011}
  (\bibinfo{year}{2007}), \eprint{astro-ph/0612381}.

\bibitem[{\citenamefont{Sahni et~al.}(2008)\citenamefont{Sahni, Shafieloo, and
  Starobinsky}}]{Sahni:2008xx}
\bibinfo{author}{\bibfnamefont{V.}~\bibnamefont{Sahni}},
  \bibinfo{author}{\bibfnamefont{A.}~\bibnamefont{Shafieloo}},
  \bibnamefont{and} \bibinfo{author}{\bibfnamefont{A.~A.}
  \bibnamefont{Starobinsky}}, \bibinfo{journal}{Phys. Rev.}
  \textbf{\bibinfo{volume}{D78}}, \bibinfo{pages}{103502}
  (\bibinfo{year}{2008}), \eprint{0807.3548}.

\bibitem[{\citenamefont{Sullivan et~al.}(2007)\citenamefont{Sullivan, Cooray,
  and Holz}}]{Sullivan:2007pd}
\bibinfo{author}{\bibfnamefont{S.}~\bibnamefont{Sullivan}},
  \bibinfo{author}{\bibfnamefont{A.}~\bibnamefont{Cooray}}, \bibnamefont{and}
  \bibinfo{author}{\bibfnamefont{D.~E.} \bibnamefont{Holz}},
  \bibinfo{journal}{JCAP} \textbf{\bibinfo{volume}{0709}}, \bibinfo{pages}{004}
  (\bibinfo{year}{2007}), \eprint{arXiv:0706.3730 [astro-ph]}.

\bibitem[{\citenamefont{Riess et~al.}(2007)}]{Riess:2006fw}
\bibinfo{author}{\bibfnamefont{A.~G.} \bibnamefont{Riess}}
  \bibnamefont{et~al.}, \bibinfo{journal}{Astrophys. J.}
  \textbf{\bibinfo{volume}{659}}, \bibinfo{pages}{98} (\bibinfo{year}{2007}),
  \eprint{astro-ph/0611572}.

\bibitem[{\citenamefont{Eisenstein et~al.}(2005)}]{Eisenstein:2005su}
\bibinfo{author}{\bibfnamefont{D.~J.} \bibnamefont{Eisenstein}}
  \bibnamefont{et~al.} (\bibinfo{collaboration}{SDSS}),
  \bibinfo{journal}{Astrophys. J.} \textbf{\bibinfo{volume}{633}},
  \bibinfo{pages}{560} (\bibinfo{year}{2005}), \eprint{astro-ph/0501171}.

\bibitem[{\citenamefont{Tegmark et~al.}(2004)}]{Tegmark:2003uf}
\bibinfo{author}{\bibfnamefont{M.}~\bibnamefont{Tegmark}} \bibnamefont{et~al.}
  (\bibinfo{collaboration}{SDSS}), \bibinfo{journal}{Astrophys. J.}
  \textbf{\bibinfo{volume}{606}}, \bibinfo{pages}{702} (\bibinfo{year}{2004}),
  \eprint{astro-ph/0310725}.

\bibitem[{\citenamefont{Spergel et~al.}(2007)}]{Spergel:2006hy}
\bibinfo{author}{\bibfnamefont{D.~N.} \bibnamefont{Spergel}}
  \bibnamefont{et~al.} (\bibinfo{collaboration}{WMAP}),
  \bibinfo{journal}{Astrophys. J. Suppl.} \textbf{\bibinfo{volume}{170}},
  \bibinfo{pages}{377} (\bibinfo{year}{2007}), \eprint{astro-ph/0603449}.

\bibitem[{\citenamefont{Huterer and Cooray}(2005)}]{Huterer:2004ch}
\bibinfo{author}{\bibfnamefont{D.}~\bibnamefont{Huterer}} \bibnamefont{and}
  \bibinfo{author}{\bibfnamefont{A.}~\bibnamefont{Cooray}},
  \bibinfo{journal}{Phys. Rev.} \textbf{\bibinfo{volume}{D71}},
  \bibinfo{pages}{023506} (\bibinfo{year}{2005}), \eprint{astro-ph/0404062}.

\bibitem[{\citenamefont{Hicken et~al.}(2009{\natexlab{a}})}]{Hicken:2009df}
\bibinfo{author}{\bibfnamefont{M.}~\bibnamefont{Hicken}} \bibnamefont{et~al.}
  (\bibinfo{year}{2009}{\natexlab{a}}), \eprint{0901.4787}.

\bibitem[{\citenamefont{Hicken et~al.}(2009{\natexlab{b}})}]{Hicken:2009dk}
\bibinfo{author}{\bibfnamefont{M.}~\bibnamefont{Hicken}} \bibnamefont{et~al.}
  (\bibinfo{year}{2009}{\natexlab{b}}), \eprint{0901.4804}.

\bibitem[{\citenamefont{Cardone et~al.}(2009)\citenamefont{Cardone,
  Capozziello, and Dainotti}}]{Cardone:2009mr}
\bibinfo{author}{\bibfnamefont{V.~F.} \bibnamefont{Cardone}},
  \bibinfo{author}{\bibfnamefont{S.}~\bibnamefont{Capozziello}},
  \bibnamefont{and} \bibinfo{author}{\bibfnamefont{M.~G.}
  \bibnamefont{Dainotti}} (\bibinfo{year}{2009}), \eprint{0901.3194}.

\bibitem[{\citenamefont{Percival et~al.}(2007)}]{Percival:2007yw}
\bibinfo{author}{\bibfnamefont{W.~J.} \bibnamefont{Percival}}
  \bibnamefont{et~al.}, \bibinfo{journal}{Mon. Not. Roy. Astron. Soc.}
  \textbf{\bibinfo{volume}{381}}, \bibinfo{pages}{1053} (\bibinfo{year}{2007}),
  \eprint{arXiv:0705.3323 [astro-ph]}.

\bibitem[{\citenamefont{Shafieloo et~al.}(2009)\citenamefont{Shafieloo, Sahni,
  and Starobinsky}}]{Shafieloo:2009ti}
\bibinfo{author}{\bibfnamefont{A.}~\bibnamefont{Shafieloo}},
  \bibinfo{author}{\bibfnamefont{V.}~\bibnamefont{Sahni}}, \bibnamefont{and}
  \bibinfo{author}{\bibfnamefont{A.~A.} \bibnamefont{Starobinsky}}
  (\bibinfo{year}{2009}), \eprint{0903.5141}.

\end{thebibliography}

\end{document}